\documentclass{iaus}
\usepackage{graphicx}

\title[Multi-wavelength probes of distant lensed galaxies] 
{Multi-wavelength probes of distant lensed galaxies}

\author[Stephen Serjeant]   
{Stephen Serjeant$^1$
}

\affiliation{$^1$Dept. of Physical Sciences, The Open University, Milton Keynes, MK7 6AA, UK}

\pubyear{2011} 
\volume{284}  
\pagerange{1--12}
\jname{The Spectral Energy Distribution of Galaxies}
\editors{R.J. Tuffs \&  C.C.Popescu, eds.}
\begin{document}

\maketitle

\begin{abstract}
  I summarise recent results on multi-wavelength properties of distant
  lensed galaxies, with a particular focus on {\it Herschel}. Submm surveys
  have already resulted in a breakthrough discovery of an extremely
  efficient selection technique for strong gravitational
  lenses. Benefitting from the gravitational magnification boost,
  blind mm-wave redshifts have been demonstrated on IRAM, SMA and GBT,
  and follow-up emission line detections have been made of water,
  [O{\sc iii}], [C{\sc ii}] and other species, revealing the
  PDR/XDR/CRDR conditions. I also discuss HST imaging of submm lenses,
  lensed galaxy reconstruction, the prospects for ALMA and e-Merlin and the
  effects of differential magnification. Many emission line
  diagnostics are relatively unaffected by differential magnification,
  but SED-based estimates of bolometric fractions in lensed infrared
  galaxies are so unreliable as to be useless, unless a lens mass
  model is available to correct for differential amplification.
\end{abstract}

\firstsection 
\section{Introduction}
Gravitational lens magnification is a fast route to faint object and
faint population studies. This technique has been exploited to
reveal or constrain the populations responsible for integrated far-IR
and submm extragalactic backgrounds (e.g. \cite[Smail et al. 1997]{Smail97},
\cite[Knudsen et al. 2008]{Knudsen08}).  Lensing is also one of the
few available means of inferring the distribution of dark matter in
galaxies (e.g. \cite[Gavazzi et al. 2007]{Gavazzi07}). Rare lens
configurations even have the promise of yielding $\Omega_{\rm M}$ and
$w$ to $10\%$; this requires $50$ double Einstein rings, requiring in
turn a parent sample of thousands of lenses (\cite[Gavazzi et
al. 2008]{Gavazzi08}). Indeed many current and future applications of
lensing are limited by sample size (\cite[Treu
2010]{Treu10}). Furthermore, most lens surveys have selected strongly
for foreground lenses at relatively low redshifts, e.g. $z<0.2$,
restricting studies of the evolving dark matter distribution in
galaxies. These are strong drivers for both larger and
higher redshift samples of lenses.

The past year has seen an explosion of interest in submm-selected
strong gravitational lenses, mainly driven by {\it Herschel} open and
guaranteed time surveys and their follow-ups. New techniques pioneered
with {\it Herschel} are capable of detecting lenses in much higher
numbers and to much higher redshifts than previous optical/near-IR
surveys such as SLACS or SLQS (\cite[Bolton et al. 2006]{Bolton06},
\cite[Oguri et al. 2006]{Oguri06}). This paper will briefly review
some recent key infrared lensing results with a particular focus on
the multiwavelength spectral energy distributions and on new data
from {\it Herschel}.

\section{Galaxy cluster lenses}
Some of the first images made public from {\it Herschel} were the
SPIRE guaranteed time $250-350-500\,\mu$m images of the galaxy cluster
lens Abell\,2218. This well-known system is one of several to have
been used to constrain the $850\,\mu$m background population
(e.g. \cite[Knudsen et al. 2008]{Knudsen08}). Recently,
\cite{Hopwood10} had also used the Japanese {\it AKARI} facility to
make its deepest $15\,\mu$m galaxy population constraints in this
field, finding they could account for $87\%$ of the predicted
extragalactic background at this wavelength. At the time of writing,
the SPIRE data in this cluster has not been published despite the data
release, but Hopwood et al. (in prep.) have stacked the {\it AKARI}
population in the submm maps and found that at least $41\%$ ($27\%$)
of the background at $500\,\mu$m ($250\,\mu$m) is attributable
directly to the $15\,\mu$m population.  At far-IR wavelengths,
\cite{Altieri10} presented {\it Herschel} guaranteed time PACS data of
this cluster at $100-160\,\mu$m, demonstrating a constraint on the
source counts comparable with that of the ultra-deep GOODS-North maps
and accounting for $(55\pm24)\%$ and $(77\pm31)\%$ of the DIRBE direct
measurements at $100\,\mu$m and $160\,\mu$m respectively. The
prospects are clearly excellent for more extensive {\it Herschel} cluster
lens work (e.g. \cite[Egami et al. 2010]{Egami10}). 

\section{Field lenses: the promise of Herschel}
\subsection{Proofs of concepts}
Gravitational lenses were originally hard to discover. For example,
the Cosmic Lens All-Sky Survey (CLASS) observed nearly 12,000
flat-spectrum radio sources, finding 16 lenses (\cite{Myers03}). 
The large public SDSS database led to more rapid lens discovery
(e.g. \cite{Bolton06}), though at low lens redshifts by virtue
of the selection criteria. 

The steep source counts in the submm imply a strong magnification
bias. Prior to {\it Herschel}, the expectation was that about $50\%$ of
galaxies with $500\,\mu$m flux densities above $S_{500}>100$\,mJy
would be lenses, with the remainder easily identifiable as local
galaxies from optical imaging or as radio-loud AGN from radio data
(\cite{Negrello07}). This prediction was supported by the mm-wave SPT
counts. Spectacular confirmation came with the first {\it Herschel}
candidates from H-ATLAS: in its first five lens candidates,
\cite{Negrello10} showed clear evidence of 2-image and 4-image lensing
configurations from SMA data while Keck imaging showed only a
foreground elliptical.  There were therefore 5 lenses from first 5
candidates! The technique is much more sensitive to higher-redshift
foreground lenses than SDSS-based selection and holds the promise of
quickly and efficiently selecting large numbers of
lenses.\footnote{The {\it Herschel} lensing was also covered by the BBC series
  Bang Goes The Theory in September 2011, in which the Open University
  are co-producers and partners.}

An associated breakthrough has been ``blind'' submm/mm-wave redshifts
(e.g. \cite{Lupu10}, \cite{Frayer11}, \cite{Scott11}) with Z-spec (CSO) and
Zpectrometer (GBT), with IRAM confirmations. This finally circumvents
the torturous process of multi-stage multi-wavelength cross-IDs and
exhaustive 8/10-m-class spectroscopic campaigns. For lensing systems,
it provides rapid and unequivocal confirmation of a submm emitter
background to the optical galaxy.

\subsection{Spectroscopic diagnostics}
The redshift determination of blind submm/mm-wave spectroscopy also
makes follow-up far-IR and submm emission line diagnostics possible.
SPIRE FTS observations of redshift $z=3.0$ H-ATLAS lens ID\,81 by
\cite{Valtchanov11} found first detection of $88\,\mu$m [O{\sc iii}]
line at $z>0.05$. The high [O{\sc iii}]/far-IR ratio and the limit on
[O{\sc i}]/[C{\sc ii}] suggests an AGN contributes ionizing radiation,
as also suggested by the high radio/far-IR ratio. \cite{Cox11} detect
multiple CO transitions and [C{\sc ii}] in the $z=4.3$ H-ATLAS galaxy
SDP\,ID\,141; their interpretation is of a PDR with a warm ($\simeq
40$\,K) dense ($10^4$\,cm$^{-3}$) gas, and the low line/far-IR ratio
suggests again a high ionization parameter. Progress in detecting
these diagnostics is fast: \cite{Lupu11} present early results from
high-density gas tracers in lensed submm galaxies, including HNC, HCN,
HCO$^+$ and $^{13}$CO.

\cite{Omont11} detected an H$_2$O transition in the $z=2.3$ H-ATLAS
galaxy SDP\,ID\,17, arguing against a PDR on the grounds that the
luminosity ratio $L({\rm H}_2{\rm O}~2_{02}-1_{01})/L({\rm CO}~(8-7)$
is comparable to that of Mrk\,231. This, plus the fact that $L({\rm
  H}_2{\rm O}~2_{02}-1_{01})/L_{\rm far-IR}$ is greater than that of
Mrk\,231, suggests the presence of an AGN. \cite{VanderWerf11}
detected four rotational H$_2$O transitions in the $z=3.9$ lensed
quasar APM\,08279+5255, inferring the presence of warm gas
($105\pm21$\,K). An immediate corollary of both detections is
that ALMA will detect H$_2$O in many more submm lenses.

\subsection{Imaging diagnostics}
Early work on the SEDs of the submm lenses found the background
sources are very difficult or impossible to detect in optical imaging,
while {\it Spitzer} $3.6-4.5\,\mu$m data can be used to detect the
background sources \cite[(Hopwood et al. 2011)]{Hopwood11}. The
immediate corollary is that {\it there is a class of lenses that are
  missed entirely in optically-selected samples}. The first HST data
on these lenses (Negrello et al. in prep.) finds the first
submm-selected Einstein rings, and with the benefit of the prior from
the HST imaging, faint features can now be discerned in the
ground-based imaging. Submm galaxies are known to have a wide variety
of colours and obscurations so one might expect many to be detectable
by {\it Euclid} at least in its near-infrared channel. The {\it
  Herschel} samples will be a crucial training set for strong lens
discovery in {\it Euclid}.

\subsection{The luminosity function and refined lens selection}
\cite{Lapi11} derived the submm luminosity function (LF) of H-ATLAS
galaxies at \mbox{$z>1$} using a far-IR/submm colour-based photometric
redshift estimator, calibrated against redshift
determinations from CO spectroscopy. Local spiral galaxies and
lensed sources were eliminating by rejecting galaxies with optical
IFs. The derived evolving LF agrees with
the predictions of an updated version of the \cite{Granato01} model. 

The steepness of the bright-end slope of the submm LF, together with
the submm K-corrections, are the underlying 
causes of the steepness of the submm source counts. However the
steepness of the LF suggests a means to improve the
efficiency of the lens selection, and to extend it to fainter submm
fluxes: if one can select galaxies at the bright end of the {\it
  luminosity function} then they should be prone to magnification bias
in a similar way to the bright submm source counts. To do this
requires relaxing the constraint in the \cite{Lapi11} analysis of
rejecting galaxies with optical IDs. Instead, \cite{Gonzalez11} use a
high-redshift selection of submm galaxies based on their far-IR/submm
colours, then use VIKING near-IR data and search for optical/submm
photometric inconsistencies. This approach has led to the discoveries
of lens candidates fainter than the $100\,$mJy $500\,\mu$m flux
threshold of \cite{Negrello10}. The source counts of their lens
candidates agree with magnification bias predictions of the
\cite{Lapi11} luminosity functions. Extrapolating to the full H-ATLAS
survey, the authors estimate an astonishing $>10^3$ strong lenses can
be reliably detected in H-ATLAS alone, several times more than
from using a simple monochromatic submm flux cut.

\section{Differential magnification: curse or blessing?}
Gravitational lensing is purely geometrical and therefore
wavelength-independent. None-theless, in any astrophysical lens the
magnification varies with position in the source plane, so if the
background source is extended with intrinsic colour gradients, then
the lensed source may have colours different to the unlensed
case. This differential magnification could have a very significant
effect on broad-band SEDs as well as on emission line
diagnostics. Many authors assume, implicitly or explicitly, that these
differential effects can be neglected, but this assumption is often
wholly unjustified. \cite[Serjeant (2011)]{Serjeant11} makes a statistical assessment
of differential magnification for various photometric and
spectroscopic diagnostics, a range of source redshifts and a constant
comoving density of lenses. Here, I present the special case of a lens
redshift ($z_{\rm l}=0.9$) and source redshift ($z_{\rm s}=2.286$)
identical to those in the lens system IRAS\,F10214+4724.

The model background source is based on the Cosmic Eyelash
(\cite{Swinbank10}), QSO\,1148+5251 (\cite{Walter09}) and Mrk\,231
(\cite{VanderWerf10}). There are three continuum components: an AGN, a
starburst and cirrus, with relative bolometric fractions of
0.3:0.5:0.2. The AGN has a torus SED from \cite{Nenkova08}
peaking in the mid-IR, modelled as a circular region with $0.1\,$kpc
radius. The starburst region is concentrated in four identical
$50$\,pc-radius GMCs as per the Cosmic Eyelash, with a starburst SED
from \cite{Efstathiou00}. The cirrus is an extended ellipse with
$2.5$\,kpc major axis, ellipticity of $0.4$ and a cirrus SED from
\cite{Efstathiou00}. There are also emission line regions: a cool CO
region tracing the cirrus with clumps modelled in the LVG
approximation with a kinetic temperature of $20\,$K, density
$100\,$cm$^{-3}$, CO column $10^{14}$\,cm$^{-2}$ and an external
background temperature of $2.73(1+z_{\rm s})\,$K; warm CO regions with
$0.4$\,kpc radii centred on the GMCs, with LVG model temperatures of
$45\,$K, densities $10^{4}\,$cm$^{-3}$ and CO column
$10^{18}\,$cm$^{-2}$; a circular warm H$_2$O region with $120\,$pc
radius and a cool H$_2$O region with major axis $1\,$kpc (both H$_2$O
regions follow the precedent of Mrk\,231); a circular [C{\sc ii}]
emission region offset by $0.6\,$kpc from the AGN, with a radius of
$0.75\,$kpc, following the precedent of QSO\,J1148+5251. The
foreground lens is modelled as the sum of a de Vaucouleurs' profile
and a Navarro-Frenk-White profile, both typical of the SLACS
population. The simulation is conducted at a resolution of $0.001''$.

\begin{figure}
\hspace*{-1cm}\includegraphics[width=3.2in]{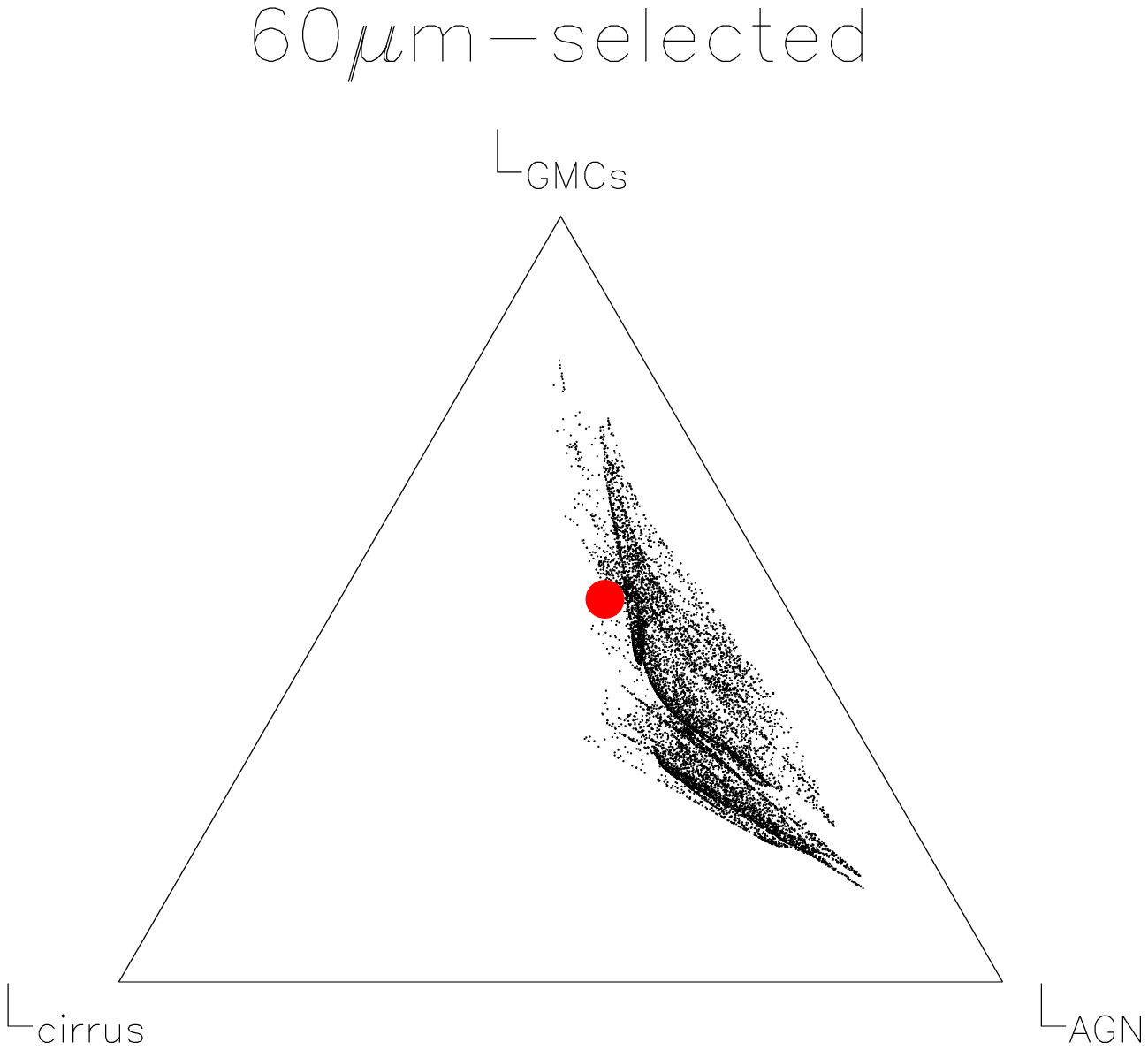} 
\hspace*{-1cm}\includegraphics[width=3.2in]{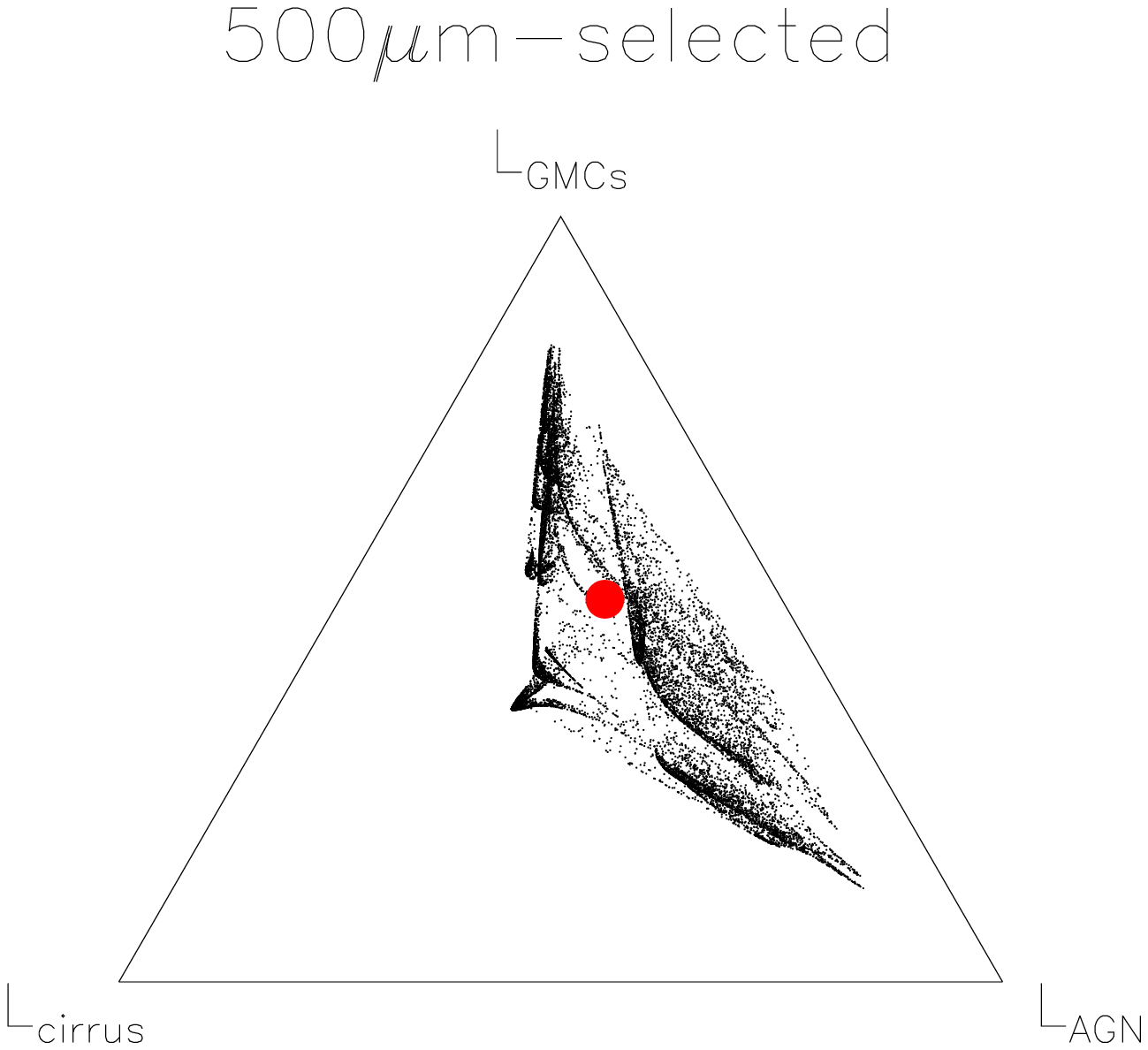}
\begin{center}
  \caption{Bolometric fractions of strongly-lensed galaxies
    (magnifications $\mu>10$, small dots, sparse-sampled for clarity)
    compared to the underlying background source fractions (large
    filled circle). These are ternary diagrams, i.e. the bolometric
    fraction is proportional to the perpendicular distance from an
    edge. At each apex, the bolometric fraction at that apex for that
    component is $100\%$, while the fraction is $0\%$ at the opposite
    edge. Note the different distributions in $500\,\mu$m-selected and
    $60\,\mu$m-selected lenses.  }
   \label{fig:ternary}
\end{center}
\end{figure}

\begin{figure}
\hspace*{-1cm}\includegraphics[width=3in]{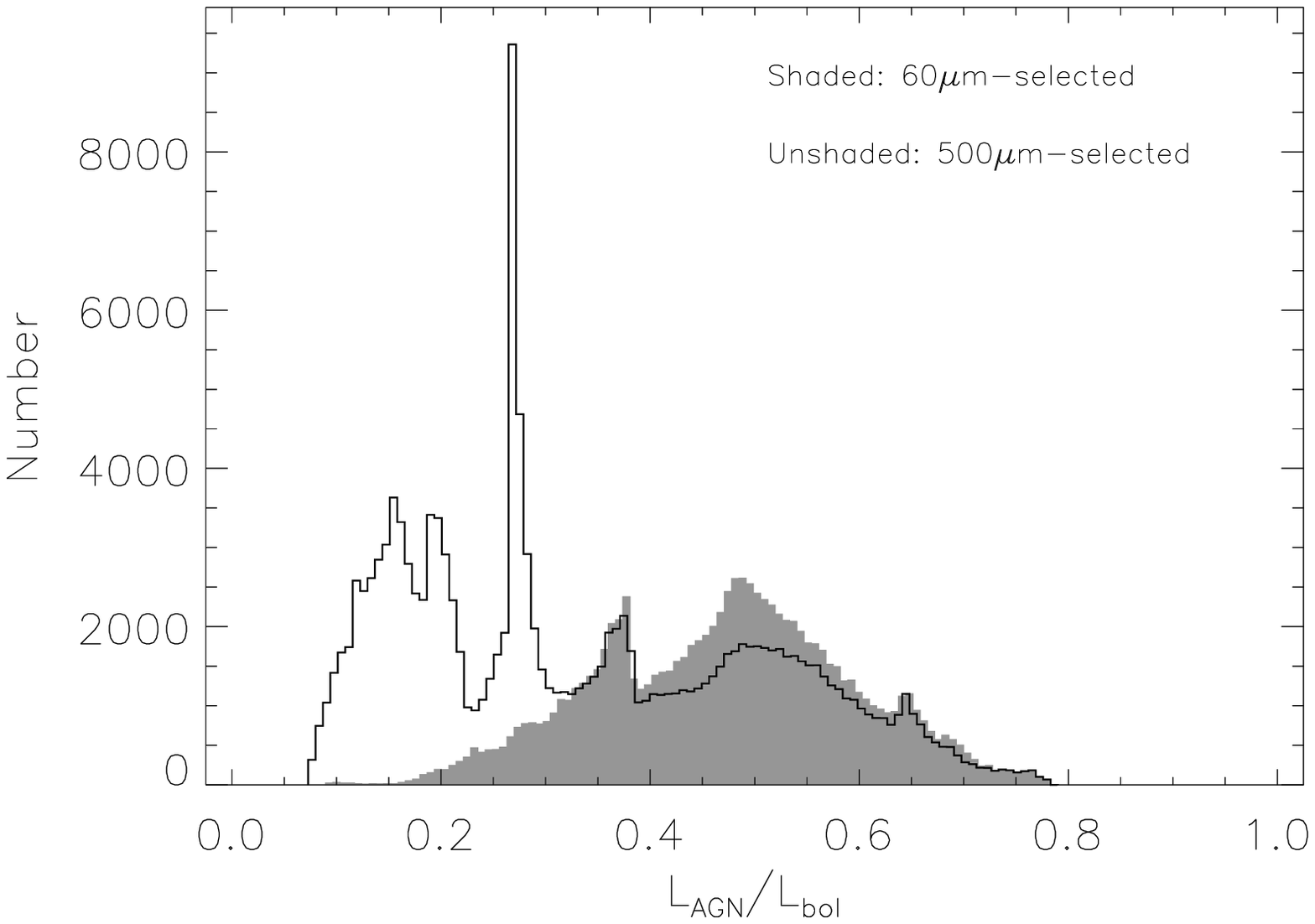} 
\hspace*{-0.5cm}\includegraphics[width=3.2in]{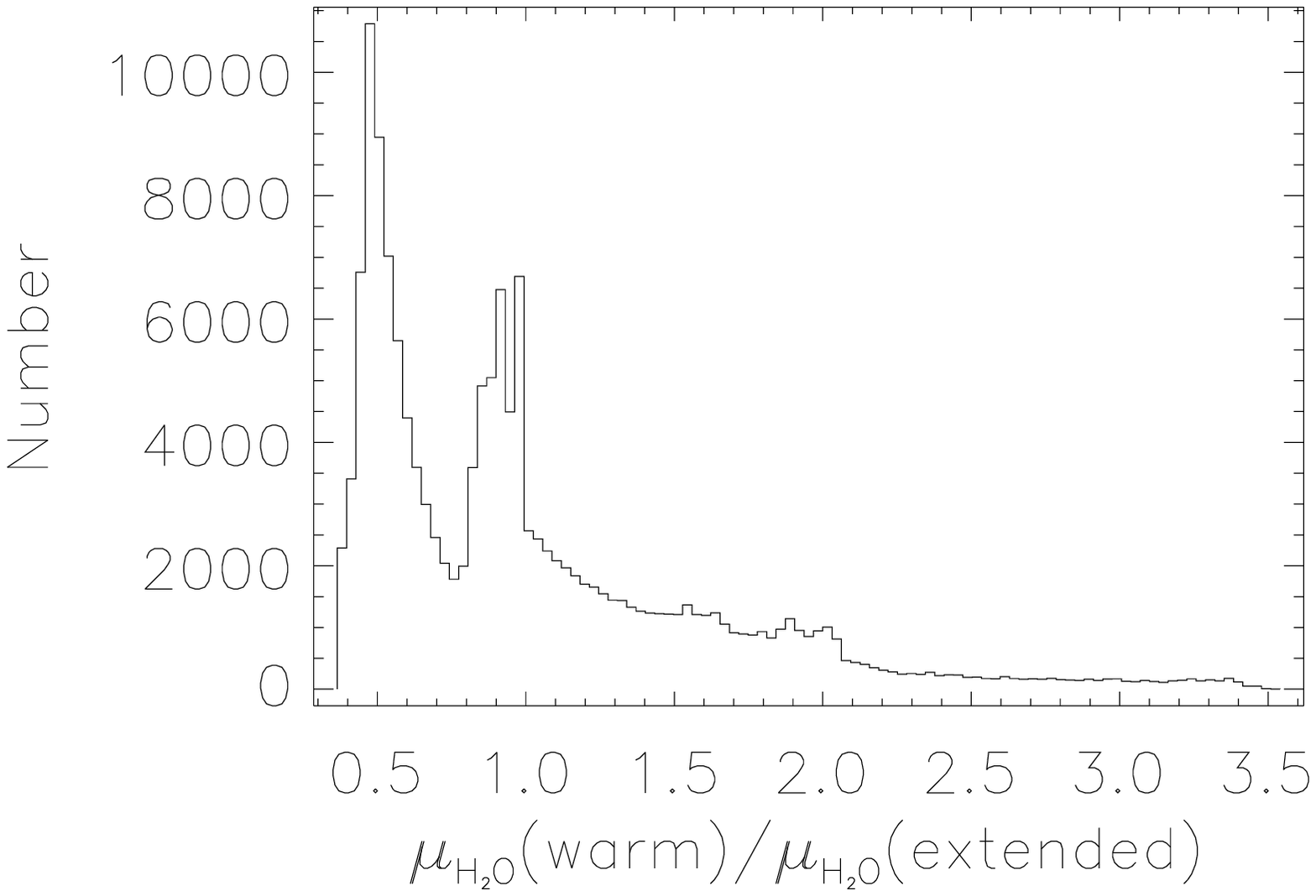}
\begin{center}
  \caption{ Left: AGN bolometric fractions in the simulated lensed
    galaxy, with monochromatic magnifications $\mu>10$. Note that the
    probability distribution of bolometric fractions depends strongly
    on the selection wavelength. Right: relative magnification factors
    of the warm and cool H$_2$O lines discussed in the text, for a
    $500\,\mu$m-selected lenses with magnification
    $\mu>10$. }
   \label{fig:h2o}
\end{center}
\end{figure}

A clear result of these simulations is that the apparent bolometric
fractions of strong lenses (magnification $\mu>10$) depend very
strongly on the selection wavelength. Fig.\,\ref{fig:ternary} shows
the bolometric fractions for a simulated lensed galaxy selected at
$60\,\mu$m and at $500\,\mu$m, while Fig.\,\ref{fig:h2o} (left) shows
the AGN bolometric fraction for the lensed galaxies selected at these
wavelengths.  At this source redshift, $60\,\mu$m (observed-frame) is
close to the peak of the AGN bolometric output, while observations at
$500\,\mu$m predominantly sample the star-forming and cirrus
components. Clearly the probability distribution of the AGN bolometric
fraction depends very strongly on the selection wavelength, though
some lens configurations are common to both wavelengths.  The effect
is much weaker for moderate-magnification systems ($2<\mu<5$); in
these systems, the observed bolometric fractions {\it are} unbiased
estimators of the underlying values.

Another striking result is that the warm and cool H$_2$O regions are
also subject to differential magnification, at least in
strongly-lensed systems (magnifications $\mu>10$). Fig.\,\ref{fig:h2o}
shows the relative magnification factors of the warm and cool H$_2$O
emission line regions. These components contribute a similar flux in
Mrk\,231 for upper energy levels under $200\,$K, while the warm
component dominates at higher temperatures. There is clearly a long
tail in Fig.\,\ref{fig:h2o} in which the warm component is boosted
relative to the cooler, more extended component. A similar tail is
present in the H$_2$O {\it vs.} $500\,\mu$m magnifications, and in
H$_2$O {\it vs.} GMCs. The presence of this tail provides an
alternative explanation for the detection of H$_2$O ($2_{02}-1_{11}$)
in an H-ATLAS galaxy (\cite{Omont11}) because $\simeq2/3$ of the flux
of this line in Mrk\,231 comes from the compact warm component. 
Having said that, at moderate magnifications ($2<\mu<5$) there are no
appreciable differential magnification effects in the H$_2$O lines. 

Differential magnification also affects the observed bolometric
fraction of the [C{\sc ii}] line (e.g. \cite{Valtchanov11}). However,
in this case the bolometric fractions vary by $\simeq 0.1-0.2$\,dex,
comparable to the measurement uncertainties. Simulating the [C{\sc
  ii}] and CO line diagnostic diagram in \cite{Valtchanov11} (not
shown), it appears that the dynamic range of the diagnostic is much
larger than that of the differential magnification effects, so this
diagnostic is only weakly affected by differential lensing. 

\begin{figure}
\hspace*{-1cm}\includegraphics[width=3.5in]{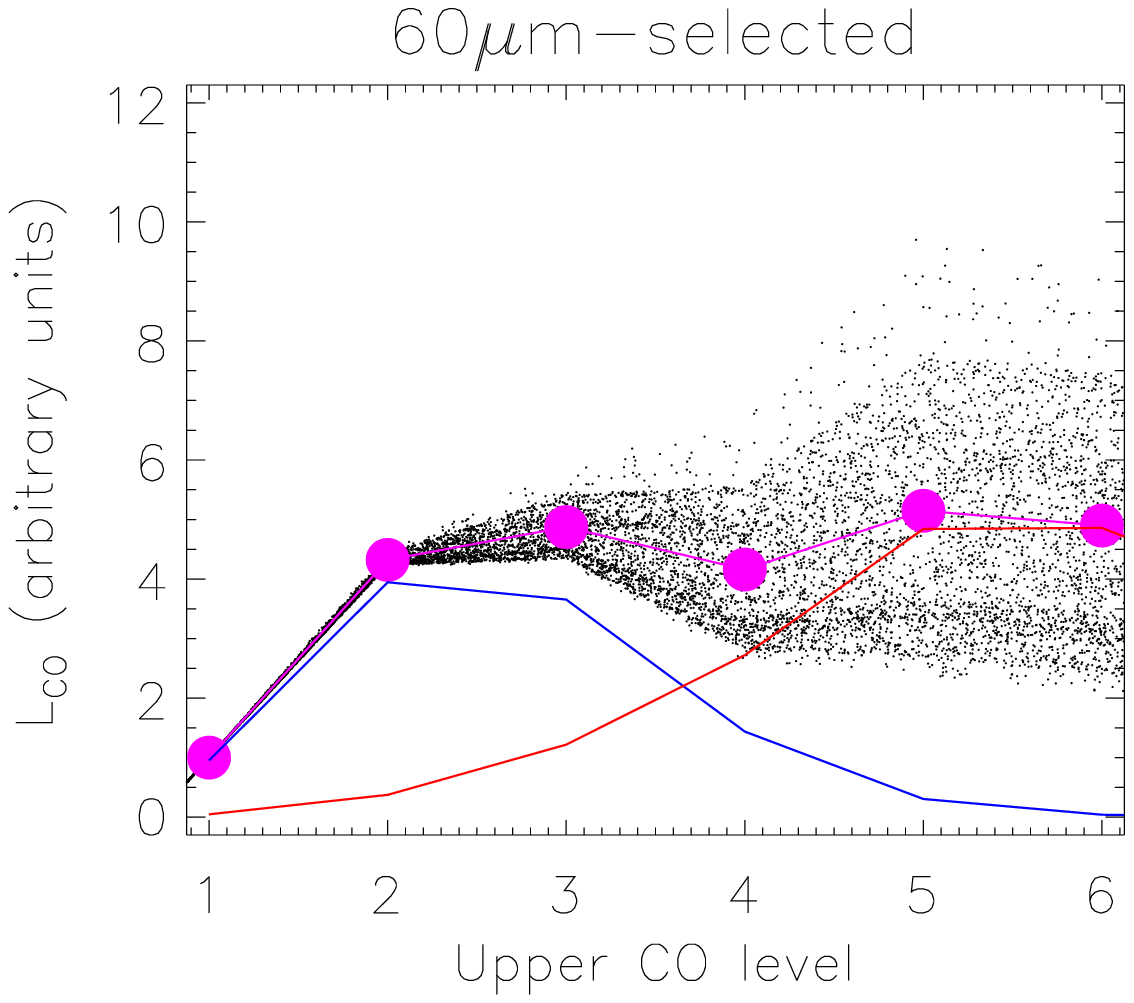} 
\hspace*{-2cm}\includegraphics[width=3.5in]{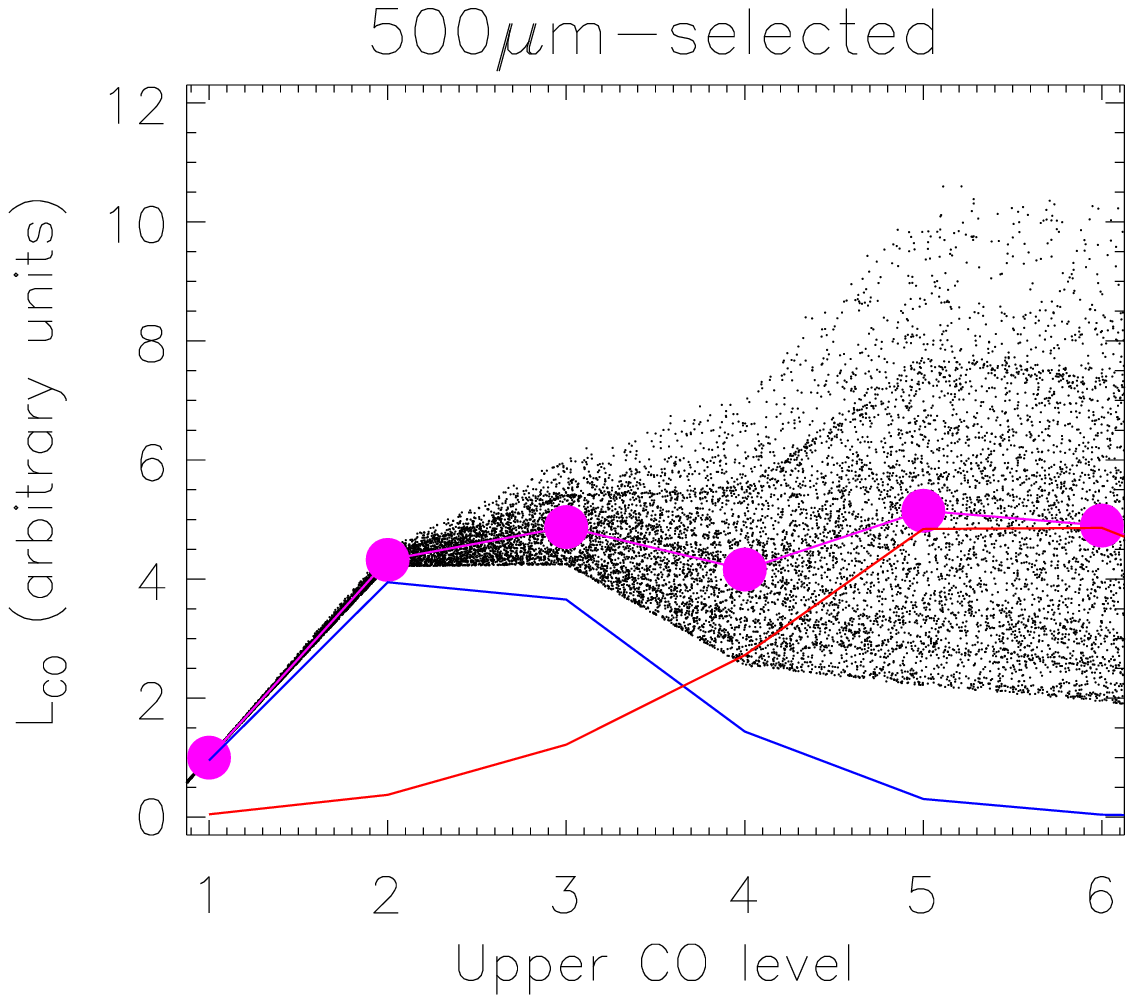}
\begin{center}
  \caption{ CO Spectral Line Energy Distribution (SLED) for the
    underlying source (large points), comprised of the warm and cool
    components (filled lines). Also shown as a shaded region is the
    range spanned by the differentially magnified simulated
    galaxies. Clearly, irrespective of selection wavelength, the CO
    SLED is strongly affected. This effect is more or less equally
    present for magnification $\mu>10$ lenses (shown) and $2<\mu<5$
    lenses.  }
   \label{fig:co}
\end{center}
\end{figure}

Finally, there is a serious distortion of the CO
ladder. Fig\,\ref{fig:co} shows the predicted lensed CO Spectral Line
Energy Distributions (SLEDs), compared to that of the underlying
background source. This time, even moderate-magnification systems are
affected. The warm CO is located very close to
the GMCs in the model galaxy, but this does not
necessarily guarantee that the warm CO is boosted relative
to the cool component, even if selecting at $500\,\mu$m. Unlike the
case of the [C{\sc ii}] and CO line diagnostic diagram in
\cite{Valtchanov11}, the useful dynamic range of the CO SLED is much
smaller, making it much more sensitive to differential magnification
effects. 

Differential magnification is therefore a serious problem if its
effects are not accounted for. Both the SED and emission line ratios
are significantly distorted in galaxies with magnifications $\mu>10$,
and the CO SLED is very strongly distorted in any system with
$\mu>2$. However if a foreground mass model is available and the
source is resolved, it can nevertheless work to our advantage. Lensing
conserves surface brightness, so regions of high magnification are
subject to angular magnification. This may, for example, provide
uniquely high angular resolution views of active nuclei or
star-forming regions in infrared-luminous galaxies.  Observations by
ALMA or eMerlin will easily detect lens morphologies, and an eMerlin
legacy survey follow-up of submm-selected lenses is already underway.

These results are robust to changes in the assumed configuration of
the background source. Only if the background source is broadly
homogeneous could strong differential magnification effects be
avoided.  One might argue that the number of GMC regions could be much
larger, e.g. to resemble those in $z=0$ spiral discs (and at variance
with observations of high-redshift infrared-luminous galaxies), but an
active nucleus will inevitably create a colour gradient in the
background source.  \cite{Serjeant11} shows that the same effects are
present for a constant comoving population of lenses, generalising the
fixed lens redshift of $z_{\rm l}=0.9$ in this paper, and for
different source redshifts.  

\section{Conclusions}
Galaxy cluster lenses are a rapid route to probing the populations
dominating the cosmic infrared backgrounds, both through direct
detections and stacking analyses. The bulk of the observed or
predicted cosmic infrared backgrounds at $15\,\mu$m, $100\,\mu$m,
$160\,\mu$m and $850\,\mu$m have now been resolved into individual
sources through ultra-deep imaging in the fields of foreground
gravitationally-lensing galaxy clusters. Meanwhile, there is a clear
need for larger numbers of strong galaxy-galaxy lens systems,
particularly at higher lens redshifts. Following the spectacular
confirmation of the $\sim100\%$ gravitational lens selection
efficiency from submm-selected lenses, there is now a tremendous scope
for such lens discovery in the submm with {\it Herschel}. A further
breakthrough has been ``blind'' submm/mm-wave redshift determination,
i.e. without recourse to the torturous multi-stage multi-wavelength
cross-identification and long-term 8m/10m-class optical/near-IR
spectroscopy. Emission line diagnostics already yielding constraints
on the physical conditions in these lensed galaxies. However,
differential magnification effects cannot be neglected. In particular,
the observed CO SLED in any lensed system can easily be
un-representative of the SLED in the underlying source. Furthermore,
the uncorrected bolometric fractions of strongly-lensed galaixes
(magnifications $\mu>10$) inferred from broad-band SEDs are so
unreliable as to be useless.

SS thanks friends and colleagues the H-ATLAS (\cite{Eales10}) and
HerMES consortia, whose results are quoted in this paper, STFC (grant
ST/G002533/1) for financial support, and the conference organisers for
the kind invitation.


\begin{thebibliography}{}
\bibitem[Altieri et al. (2010)]{Altieri10} Altieri, B., et al., 2010,
  \textit{A\&A}, 518, L17
\bibitem[Bolton et al. 2006]{Bolton06} Bolton, A.S., et al., 2006, \textit{ApJ}, 638, 703
\bibitem[Cox et al. (2011)]{Cox11}
{Cox, P., et al.}, 2011, \textit{ApJ} in press (arXiv:1107.2924)
\bibitem[Eales et al. 2010]{Eales10} Eales, S.A., et al., 2010, PASP, 122, 499
\bibitem[Efstathiou et al. (2000)]{Efstathiou00} Efstathiou, A.,
  Rowan-Robinson, M., \& Siebenmorgen, R., 2000, MNRAS, 313, 734
\bibitem[Egami et al. (2010)]{Egami10} Egami, E., et al., 2010,
  \textit{A\&A}, 518, L12
\bibitem[Frayer et al. 2011]{Frayer11} Frayer, D.T., et al., 2011, ApJL, 726, 22
\bibitem[Gavazzi et al. (2007)]{Gavazzi07} Gavazzi, R., et al., 2007,
  \textit{ApJ}, 667, 176
\bibitem[Gavazzi et al. (2008)]{Gavazzi08} Gavazzi, R., et al., 2008,
  \textit{ApJ}, 677, 1046
\bibitem[Gonz\'alez-Nuevo et al. (2011)]{Gonzalez11} Gonz\'{a}lez-Nuevo, J., et
  al., 2011, \textit{ApJ}, submitted
\bibitem[Granato et al. 2001]{Granato01} Granato, G.L., et al., 2001, MNRAS, 324, 757
\bibitem[Hopwood et al. (2010)]{Hopwood10} Hopwood, R., et al., 2010,
  \textit{ApJL}, 728, 4
\bibitem[Hopwood et al. (2011)]{Hopwood11}
{Hopwood, R., et al.}, 2011, \textit{ApJL}, 728, 4
\bibitem[Knudsen et al. (2008)]{Knudsen08} Knudsen, K.K., et al., 2008, \textit{MNRAS}, 384, 1161
\bibitem[Lapi et al. 2011]{Lapi11} Lapi, A., et al., 2011,
  \textit{ApJ}, in press (arXiv:1108.3911) 
\bibitem[Lupu et al. 2010]{Lupu10} Lupu, R., et al., 2010, ApJ submitted (arXiv:1009.5983)
\bibitem[Lupu et al. 2011]{Lupu11} Lupu, R., et al., 2011, in The Molecular Universe,
  Proceedings of the 280th Symposium of the International Astronomical
  Union, Toledo, Spain, May 30-June 3, 2011
\bibitem[Myers et al. 2003]{Myers03} Myers, S.T., et al., 2003, MNRAS, 341, 1
\bibitem[Negrello et al. 2007]{Negrello07} Negrello, M., et al., 2007, MNRAS, 377, 1557
\bibitem[Negrello et al. (2010)]{Negrello10}
{Negrello, M., et al.}, 2010, \textit{Science}, 330, 800
\bibitem[Nenkova et al. 2008]{Nenkova08} Nenkova, M., Sirocky, M.M., Ivesi$\acute{\rm c}$,
  $\breve{\rm Z}$., Elitzur, M, 2008,
  ApJ, 685, 147
\bibitem[Oguri et al. (2006)]{Oguri06} Oguri, M., et al., 2006,
  \textit{AJ}, 132, 999
\bibitem[Omont et al. 2011]{Omont11}
{Omont, A., et al.}, 2011, \textit{A\&A}, 530, L3O
\bibitem[Scott et al. 2011]{Scott11} Scott, K.S., et al., 2011, ApJ, 733, 29
\bibitem[Serjeant (2011)]{Serjeant11} Serjeant, S., 2011, MNRAS submitted
\bibitem[Smail et al. (1997)]{Smail97} Smail, I., et al., 1997, \textit{ApJL}, 490, L5
\bibitem[Swinbank et al. 2010]{Swinbank10} Swinbank, A.M., et al., 2010, Nature, 464, 733
\bibitem[Treu (2010)]{Treu10} Treu, T., 2010, \textit{ARA\&A}, 48, 87
\bibitem[Walter et al. 2009]{Walter09} Walter, F., et al., 2009, Nature, 457, 699
\bibitem[Valtchanov et al. 2011]{Valtchanov11} 
Valtchanov, I., et al., 2011, \textit{MNRAS}, in press  (arXiv:1105.3929) 
\bibitem[Van der Werf et al. 2010]{VanderWerf10} Van der Werf, P.P., et al., 2010, A\&A, 518, L42
\bibitem[Van der Werf et al. (2011)]{VanderWerf11} Van der Werf, P.P.,
  et al,. 2011, ApJ, 741, L38
%
%
%

\end{thebibliography}
\end{document}